\documentstyle[refs]{paper}
\def\vcb{\mid V_{cb} \mid}
\def\vtd{\mid V_{td} \mid}
\def\vub{\mid V_{ub}/V_{cb} \mid}
\def\f{\frac}

\def\kpnn{K^+\rightarrow\pi^+\nu\bar\nu }
\def\kpn{K^+\rightarrow\pi^+\nu\bar\nu}
\def\klpnn{K_L\rightarrow\pi^0\nu\bar\nu}
\def\klpn{K_L\rightarrow\pi^0\nu\bar\nu}

\newcommand{\be}{\begin{equation}}
\newcommand{\ee}{\end{equation}}
\newcommand{\ra}{\rightarrow}

\newcommand{\Bsg}{$B \ra X_s \gamma$ }

\newcommand{\bea}{\begin{eqnarray}}
\newcommand{\eea}{\end{eqnarray}}
\newcommand{\bd}{\begin{displaymath}}
\newcommand{\ed}{\end{displaymath}}

\newcommand{\kpiee}{$K_L \ra \pi^0 e^+ e^-$ }
\newcommand{\Lms}{\Lambda_{\overline{\rm MS}}}

\begin{document}

\title{THEORETICAL PROGRESS IN K AND B DECAYS}

\author{ Andrzej J. Buras\\
\\
\it Technische Universit\"at M\"unchen, Physik Department,\\
 D-85748 Garching, GERMANY; \\
\
 Max-Planck-Institut f\"ur Physik,\\
 -- Werner-Heisenberg-Institut --\\
 D-80805 M\"unchen, GERMANY}
\maketitle
\section*{Abstract}

We review several aspects of the recent theoretical progress
in K and B decays
including the impact of the top quark
discovery on rare and CP violating decays.
In particular we summarize the present status
of next-to-leading QCD calculations in this field stressing their
importance in the determination of the parameters
in the Cabibbo-Kobayashi-Maskawa matrix.

\section{Introduction}

An important target of particle physics is the determination
 of the unitary $3\times 3$ Cabibbo-Kobayashi-Maskawa
matrix \cite{CAB,KM} which parametrizes the charged current interactions of
 quarks:
\begin{equation}\label{1j}
J^{cc}_{\mu}=(\bar u,\bar c,\bar t)_L\gamma_{\mu}
\left(\begin{array}{ccc}
V_{ud}&V_{us}&V_{ub}\\
V_{cd}&V_{cs}&V_{cb}\\
V_{td}&V_{ts}&V_{tb}
\end{array}\right)
\left(\begin{array}{c}
d \\ s \\ b
\end{array}\right)_L
\end{equation}
The CP violation in the standard model is supposed to arise
from a single phase in this matrix.
It is customary these days to express the CKM-matrix in
terms of four Wolfenstein parameters
\cite{WO} $(\lambda,A,\varrho,\eta)$
with $\lambda=\mid V_{us}\mid=0.22 $ playing the role of an expansion
parameter and $\eta$
representing the CP violating phase:
\begin{equation}\label{2.75}
V_{CKM}=
\left(\begin{array}{ccc}
1-{\lambda^2\over 2}&\lambda&A\lambda^3(\varrho-i\eta)\\ -\lambda&
1-{\lambda^2\over 2}&A\lambda^2\\ A\lambda^3(1-\varrho-i\eta)&-A\lambda^2&
1\end{array}\right)
+O(\lambda^4)
\end{equation}
Following \cite{BLO} one can define the parameters
$(\lambda, A, \varrho, \eta)$ through
\be\label{wop}
s_{12}\equiv\lambda \qquad s_{23}\equiv A \lambda^2 \qquad
s_{13} e^{-i\delta}\equiv A \lambda^3 (\varrho-i \eta)      \ee
where $s_{ij}$ and $\delta$ enter the standard exact
parametrization \cite{PDG}  of the CKM
matrix. This specifies the higher orders terms in (\ref{2.75}).
With the definitions in
(\ref{wop}),
\begin{equation}\label{CKM1}
V_{us}=\lambda \qquad V_{cb}=A\lambda^2
\end{equation}
\begin{equation}\label{CKM2}
V_{ub}=A\lambda^3(\varrho-i\eta)
\qquad
V_{td}=A\lambda^3(1-\bar\varrho-i\bar\eta)
\end{equation}
where
\begin{equation}\label{3}
\bar\varrho=\varrho (1-\frac{\lambda^2}{2})
\qquad
\bar\eta=\eta (1-\frac{\lambda^2}{2})
\end{equation}
turn out \cite{BLO} to be excellent approximations to the
exact expressions.

\vspace{6.2cm}
\centerline{Fig. 1}

A useful geometrical representation of the CKM matrix is the unitarity
triangle obtained by using the unitarity relation
$V_{ud}V_{ub}^* + V_{cd}V_{cb}^* + V_{td}V_{tb}^* =0,$
rescaling it by $\mid V_{cd}V_{cb}^\ast\mid=A \lambda^3$ and depicting
the result in the complex $(\bar\rho,\bar\eta)$ plane as shown
in fig. 1. The lenghts CB, CA and BA are equal respectively to 1,
\begin{equation}\label{2.94a}
R_b \equiv  \sqrt{\bar\varrho^2 +\bar\eta^2}
= (1-\frac{\lambda^2}{2})\frac{1}{\lambda}
\left| \frac{V_{ub}}{V_{cb}} \right|
\qquad
{\rm and}
\qquad
R_t \equiv \sqrt{(1-\bar\varrho)^2 +\bar\eta^2}
=\frac{1}{\lambda} \left| \frac{V_{td}}{V_{cb}} \right|.
\end{equation}

The triangle in fig. 1,
 $\mid V_{us}\mid$ and $\mid V_{cb}\mid$
give the full description of the CKM matrix.
Looking at  $R_b$ and $R_t$ we observe that within
the standard model the measurements of four CP
{\it conserving } decays sensitive to $\mid V_{us}\mid$, $\mid V_{ub}\mid$,
$\mid V_{cb}\mid $ and $\mid V_{td}\mid$ can tell us whether CP violation
($\eta \not = 0 $) is predicted in the standard model.
This is a very remarkable property of
the Kobayashi-Maskawa picture of CP violation: quark mixing and CP violation
are closely related to each other.

There is of course the very important question whether the KM picture
of CP violation is correct and more generally whether the standard
model offers a correct description of weak decays of hadrons. In order
to answer these important questions it is essential to calculate as
many branching ratios as possible, measure them experimentally and
check if they all can be described by the same set of the parameters
$(\lambda,A,\varrho,\eta)$. In the language of the unitarity triangle
this means that  various curves in the $(\bar\varrho,\bar\eta)$ plane
extracted from different decays should cross each other at a single point
which determines the apex of the unitarity triangle in fig. 1.
Moreover the angles $(\alpha,\beta,\gamma)$ in the
resulting triangle should agree with those extracted one day from
CP-asymmetries in B-decays.

There is a common belief that during the coming fifteen years we will
certainly witness a dramatic
improvement in the determination of the CKM-parameters.
To this end, however, it is essential not only to perform difficult
experiments but also to have accurate formulae which would allow
a confident and  precise extraction of the CKM-parameters from the
existing and future data. We will review what
progress has been done in this direction.

Clearly the discovery of the top quark
\cite{CDF,D0} and its mass measurement had an important impact on
the field of rare decays and CP violation reducing considerably one
potential uncertainty. In loop induced K and B
decays the relevant mass parameter is the running current quark mass.
With the pole mass measurement of CDF, $m^{pole}_t=176\pm 13~GeV$,
one has $m_t^*=\bar m_t(m_t)\approx 168\pm 13~GeV$. Similarly the
D0 value $m^{pole}_t=199\pm 30~GeV$ corresponds to
$m_t^*\approx 190\pm 30~GeV$.
 In this review we will
simply denote $m_t^*$ by $m_t$.
\section{Basic Framework}
\subsection{OPE and Renormalization Group}
The basic framework for weak decays of hadrons containing u, d, s, c and
b quarks consists of the
Operator Product Expansion (OPE) combined with the renormalization group
techniques.
In this framework the amplitude for a decay $M\to F$ is written as
\begin{equation}\label{OPE}
 A(M \to F) = \langle F \mid {\cal H}_{eff} \mid M \rangle=
\frac{G_F}{\sqrt 2} V_{CKM} \sum_i
   C_i (\mu) \langle F \mid Q_i (\mu) \mid M \rangle
\end{equation}
where $ {\cal H}_{eff}$ is an effective hamiltonian relevant
for a given decay, $M$ stands for the decaying meson,
$F$ for a given final state and
$V_{CKM}$ denotes the relevant $CKM$ factor.
$ Q_i(\mu) $ denote
the local operators generated by QCD and electroweak interactions.
$ C_i(\mu) $ stand for the Wilson
coefficient functions.
The scale $ \mu $ separates the physics contributions in the ``short
distance'' contributions (corresponding to scales higher than $\mu $)
contained in $ C_i(\mu) $ and the ``long distance'' contributions
(scales lower than $ \mu $) contained in $ < F \mid Q_i (\mu) \mid M > $.
 Since  physical amplitudes
 cannot depend on $ \mu $,
the $ \mu $-dependence of $C_i(\mu)$
must be cancelled by the one present in $\langle  Q_i (\mu)\rangle $.
 It should be
stressed, however, that this cancellation generally involves many
operators due to the operator mixing under renormalization.

The $\mu$ dependence of $ C_i(\mu) $ is given by:
\begin{equation}
 \vec C(\mu) = \hat U(\mu,M_W) \vec C(M_W)
\end{equation}
where $ \vec C $ is a column vector built out of $ C_i $'s.
$\vec C(M_W)$ are the initial conditions which depend on the
short distance physics at high energy scales.
In particular they depend on $m_t$.
$ \hat U(\mu,M_W) $, the renormalization group evolution matrix,
is given as follows
\begin{equation}\label{UM}
 \hat U(\mu,M_W) = T_g exp \lbrack
   \int_{g(M_W)}^{g(\mu)}{dg' \frac{\hat\gamma^T(g')}{\beta(g')}}\rbrack
\end{equation}
with $g$ denoting QCD effective coupling constant. $ \beta(g) $
governs the evolution of $g$ and $ \hat\gamma $ is the anomalous dimension
matrix of the operators involved. The structure of this equation
makes it clear that the renormalization group approach goes
 beyond the usual perturbation theory.
Indeed $ \hat  U(\mu,M_W) $ sums automatically large logarithms
$ \log M_W/\mu $ which appear for $ \mu<<M_W $. In the so called leading
logarithmic approximation (LO) terms $ (g^2\log M_W/\mu)^n $ are summed.
The next-to-leading logarithmic correction (NLO) to this result involves
summation of terms $ (g^2)^n (\log M_W/\mu)^{n-1} $ and so on.
This hierarchic structure gives the renormalization group improved
perturbation theory. For instance
in the case of a single
operator one has including NLO corrections:
\begin{equation}\label{UMNLO}
 U (\mu, M_W) = \Biggl\lbrack 1 + {{\alpha_{QCD} (\mu)}\over{4\pi}} J
\Biggl\rbrack \Biggl\lbrack {{\alpha_{QCD} (M_W)}\over{\alpha_{QCD} (\mu)}}
\Biggl\rbrack^P \Biggl\lbrack 1 - {{\alpha_{QCD} (M_W)}\over{4\pi}} J
\Biggl\rbrack
\end{equation}
where $P$ and $J$ are given in terms of the coefficients in
the perturbative expansions for $\gamma(g)$ and $\beta(g)$.
General
formulae for $ \hat U (\mu, M_W) $ in the case of operator mixing and
valid also for electroweak effects can be found in ref.\cite{BJLW}.
The leading
logarithmic approximation corresponds to setting $ J = 0 $ in (\ref{UMNLO}).
\subsection{Classification of Operators}
The most important operators are given as follows:

{\bf Current--Current:}
\begin{equation}
Q_1 = (\bar s_{\alpha} u_{\beta})_{V-A}\;(\bar u_{\beta} d_{\alpha})_{V-A}
{}~~~~~~Q_2 = (\bar s u)_{V-A}\;(\bar u d)_{V-A}
\end{equation}

{\bf QCD--Penguins:}
\begin{equation}
Q_3 = (\bar s d)_{V-A}\sum_{q=u,d,s}(\bar qq)_{V-A}~~~~~~
 Q_4 = (\bar s_{\alpha} d_{\beta})_{V-A}\sum_{q=u,d,s}(\bar q_{\beta}
       q_{\alpha})_{V-A}
\end{equation}
\begin{equation}
 Q_5 = (\bar sd)_{V-A} \sum_{q=u,d,s}(\bar qq)_{V+A}~~~~~
 Q_6 = (\bar s_{\alpha} d_{\beta})_{V-A}\sum_{q=u,d,s}
       (\bar q_{\beta} q_{\alpha})_{V+A}
\end{equation}

{\bf Electroweak--Penguins:}
\begin{equation}
Q_7 = {3\over 2}\;(\bar s d)_{V-A}\sum_{q=u,d,s}e_q\;(\bar qq)_{V+A}
{}~~~~~ Q_8 = {3\over2}\;(\bar s_{\alpha} d_{\beta})_{V-A}\sum_{q=u,d,s}e_q
        (\bar q_{\beta} q_{\alpha})_{V+A}
\end{equation}
\begin{equation}
 Q_9 = {3\over 2}\;(\bar s d)_{V-A}\sum_{q=u,d,s}e_q(\bar q q)_{V-A}
{}~~~~~Q_{10} ={3\over 2}\;(\bar s_{\alpha} d_{\beta})_{V-A}\sum_{q=u,d,s}e_q\;
       (\bar q_{\beta}q_{\alpha})_{V-A}
\end{equation}

{\bf Magnetic--Penguins:}
\begin{equation}
Q_{7\gamma}  =  \f{e}{8\pi^2} m_b \bar{s}_\alpha \sigma^{\mu\nu}
          (1+\gamma_5) b_\alpha F_{\mu\nu}\qquad
Q_{8G}     =  \f{g}{8\pi^2} m_b \bar{s}_\alpha \sigma^{\mu\nu}
   (1+\gamma_5)T^a_{\alpha\beta} b_\beta G^a_{\mu\nu}
\end{equation}

{\bf $\Delta S = 2 $ and $ \Delta B = 2 $ Operators:}
\begin{equation}
Q(\Delta S = 2)  = (\bar s d)_{V-A} (\bar s d)_{V-A}~~~~~
 Q(\Delta B = 2)  = (\bar b d)_{V-A} (\bar b d)_{V-A}
\end{equation}

{\bf Semi--Leptonic Operators:}
\begin{equation}\label{9V}
Q_{9V}  = (\bar b s )_{V-A} (\bar e e)_{V}~~~~~
Q_{10A}  = (\bar b s )_{V-A} (\bar e e)_{A}
\end{equation}
\begin{equation}
Q(\nu\bar\nu)  = (\bar s d)_{V-A} (\nu\bar\nu)_{V-A}~~~~~
Q(\mu\bar\mu)  = (\bar s d)_{V-A} (\mu\bar\mu)_{V-A}
\end{equation}

\subsection{Towards Phenomenology}
The rather formal expression for the decay amplitudes given in
(\ref{OPE}) can always be cast in a more useful form \cite{PBE}:
\begin{equation}\label{PBEE}
A(M\to F)=\sum_i B_i V_{CKM}^{i} \eta^{i}_{QCD} F_i(m_t,m_c)
\end{equation}
In writing (\ref{PBEE})
we have generalized (\ref{OPE}) to include several CKM factors.
$F_i(m_t,m_c)$, the Inami-Lim functions,
 result from the evaluation of loop diagrams with
internal top and charm exchanges  and may also depend
solely on $m_t$ or $m_c$. In the case of current-current operators
$F_i$ are mass independent. The factors $\eta^{i}_{QCD}$ summarize
short distance QCD corrections which can be calculated by formal methods
discussed above. Finally $B_i$ stand for nonperturbative factors
related to the hadronic matrix elements of the contributing
operators: the main theoretical uncertainty in the whole enterprise.
In semi-leptonic decays such as $K\to \pi\nu\bar\nu$,
the non-perturbative $B$-factors can fortunately be determined from
leading tree level decays such as $K^+\to \pi^0 e^+\nu$ reducing
or removing the non-perurbative uncertainty. In non-leptonic
decays this is generally not possible and we have to rely on
existing non-perturbative methods. A well known example of a
$B_i$-factor is the renormalization group invariant parameter
$B_K$ \cite{BSS} defined by
\begin{equation}\label{bk}
B_K=B_K(\mu)\left[\alpha_s(\mu)\right]^{-2/9}
\qquad
\langle \bar K^{o}\mid Q(\Delta S=2)\mid K^{o}\rangle=
\frac{8}{3} B_K(\mu)F_K^2 m_K^2
\end{equation}
\subsection{Inclusive Decays}
Sofar we have discussed only {\it exclusive} decays. During the
recent years considerable progress has been made for inclusive
decays of heavy mesons. The starting point is again the effective
hamiltonian in (\ref{OPE}) which includes the short distance QCD
effects in $C_i(\mu)$. The actual decay described by the operators
$Q_i$ is then calculated in the spectator model corrected for
additional virtual and real gluon corrections.
Support for this approximation
comes from the $1/m_b $ expansions.
Indeed the spectator
model has been shown to correspond to the leading order approximation
in the $1/m_b$ expansion.
The next corrections appear at the ${\cal O}(1/m_b^2)$
level. The latter terms have been studied by several authors
\cite{Chay,Bj,Bigi} with the result that they affect various
branching ratios by less than $10\%$ and often by only a few percent.
There is a vast literature on this subject and I can only refer here to
recent reviews \cite{Bigi,Mannel} where further references can be found.
Of particular importance for this field was also the issue of the
renormalons which are nicely discussed in \cite{Beneke,Brown}.
\section{Theoretical Progress in K and B Decays}
It is impossible to review adequately the full theoretical progress
here. Let me then list only a few achievements of the last five
years which in my opinion should be considered as important
contributions to the field of weak decays.
\begin{itemize}
\item
Calculation of NLO corrections to the Wilson coefficients for
nearly all decays (ordinary, rare and CP- violating) \cite{BBL}.
\item
Applications of heavy quark effective theory to exclusive decays
which resulted in particular
in an improved determination of $V_{cb}$ \cite{Neubert}.
\item
Heavy Quark Expansions for inclusive decays (see reviews in
\cite{Bigi,Mannel}),
which by putting the spectator model on a firmer ground allow
for an improved determination of $V_{cb}$ in agreement with the
exclusive determination \cite{Bigi,Brown}.
\item
Some progress in the calculations of non-perturbative parameters
such as $B_K$ and $F_B$.
\item
Identification of  decays  nearly without any hadronic
uncertainties.
\end{itemize}
In this review I will mainly discuss the first and the last item
on this list, incorporating in this discussion the achievements related
to the remaining three items.
\section{Weak Decays Beyond Leading Logarithms}
\subsection{General Remarks}
Until 1989 all the calculations in the field of weak
decays were done in the leading logarithmic approximation
except for \cite{ALTA}
where NLO QCD corrections to the Wilson
coefficients of the current-current operators have been calculated.
Today the effective hamiltonians for weak decays are
available at the next-to-leading level for the most important
and interesting cases due to a series of publications
listed in
table 1. We will discuss this list briefly below.
An extended version of this discussion appeared recently \cite{ABU95}.
  A very detailed review
of the existing NLO calculations will appear soon \cite{BBL}.

Let us recall why NLO calculations are important for the
phenomenology of weak decays.

\begin{itemize}
\item The NLO is first of all necessary to test the validity of
the renormalization group improved perturbation theory.
\item Without going to NLO the QCD scale $\Lambda_{\overline{MS}}$
extracted from various high energy processes cannot be used
meaningfully in weak decays.
\item Due to renormalization group invariance the physical
amplitudes do not depend on the scales $\mu$ present in $\alpha_s$
or in the running quark masses, in particular $m_t(\mu)$,
$m_b(\mu)$ and $m_c(\mu)$. However
in perturbation theory this property is broken through the truncation
of the perturbative series. Consequently one finds sizable scale
ambiguities in the leading order, which can be reduced considerably
by going to NLO.
\item The central issue of the top quark mass dependence
is often a NLO effect.
\end{itemize}

\begin{table}
\begin{center}
\begin{tabular}{|l|l|}
\hline
\bf \phantom{XXXXXX} Decay & \bf \phantom{XX} Reference~~~ \\
\hline
\hline
\multicolumn{2}{|c|}{$\Delta F=1$ Decays} \\
\hline
current-current operators     & \cite{ALTA,BW} \\
QCD penguin operators         & \cite{BJLW1,BJLW,ROMA1,ROMA2} \\
electroweak penguin operators & \cite{BJLW2,BJLW,ROMA1,ROMA2} \\
magnetic penguin operators    & \cite{MisMu:94}  \\
$Br(B)_{SL}$                  & \cite{ALTA,Buch:93,Bagan} \\
\hline
\multicolumn{2}{|c|}{Particle-Antiparticle Mixing} \\
\hline
$\eta_1$                   & \cite{HNa} \\
$\eta_2,~\eta_B$           & \cite{BJW} \\
$\eta_3$                   & \cite{HNb} \\
\hline
\multicolumn{2}{|c|}{Rare K- and B-Meson Decays} \\
\hline
$K^0_L \rightarrow \pi^0\nu\bar{\nu}$, $B \rightarrow l^+l^-$,
$B \rightarrow X_{\rm s}\nu\bar{\nu}$ & \cite{BB1,BB2} \\
$K^+   \rightarrow \pi^+\nu\bar{\nu}$, $K_L \rightarrow \mu^+\mu^-$
                                      & \cite{BB3} \\
$K^+\to\pi^+\mu\bar\mu$               & \cite{BB5} \\
$K_L \rightarrow \pi^0e^+e^-$         & \cite{BLMM} \\
$B\rightarrow X_s e^+e^-$           & \cite{Mis:94,BuMu:94} \\
\hline
\end{tabular}
\end{center}
\centerline{}
\caption{References to NLO Calculations}
\end{table}

\subsection{Current-Current Operators}
The NLO corrections to the coefficients of $Q_1$ and $Q_2$ have been
first calculated by Altarelli et al.\cite{ALTA}
 using the Dimension Reduction
Scheme (DRED) for $\gamma_5$. In 1989 these coefficients have been
calculated in DRED, NDR and HV schemes for $\gamma_5$ by Peter Weisz and
myself \cite{BW}. The result for DRED obtained by the Italian group
has been
confirmed. The coefficients $C_1$ and $C_2$ show a rather strong
renormalization scheme dependence which in physical quantities
should be cancelled by the one present in the matrix elements of
$Q_1$ and $Q_2$. This cancellation has been shown explicitly in
\cite{BW} demonstrating thereby the compatibilty of the results for
$C_1$ and $C_2$ in DRED, NDR and HV schemes. A recent discussion
of $C_1(\mu)$ and $C_2(\mu)$ in these schemes can be found in
\cite{AB:95c}.
\subsection{NLO Corrections to $B_{SL}$}
A direct physical application of the NLO corrections to $C_1$ and
$C_2$ is the calculation of the non-leptonic width
for B-Mesons which is relevant for the theoretical prediction of
the inclusive semileptonic branching ratio $B_{SL}$
in B-decays.
This calculation can be done within the spectator model corrected
for small non-perturbative corrections \cite{Bigi}
 and more important gluon
bremsstrahlung and virtual gluon corrections.
The calculation of $B_{SL}$
for massless final quarks has been done by Altarelli et al.\cite{ALTA}
 in the
DRED scheme and by Buchalla \cite{Buch:93} in the HV scheme.
 The results of these papers agree with each other.

Unfortunately the theoretical branching ratio
based on the QCD calculation of refs. \cite{ALTA,Buch:93}  give typically
$B_{SL}=12.5-13.5\%$ \cite{AP:92}
whereas the experimental world average \cite{PDG} is
$B^{exp}_{SL}=(10.43\pm 0.24)\%$.
The inclusion of the leading non-perturbative correction
${\cal O}(1/m_b^2)$ lowers slightly the theoretical
prediction but gives only $\Delta_{NP} B_{SL}=-0.2\%$ \cite{Bigi}.
On the other hand
Bagan et al. \cite{Bagan}
have demonstrated that including
mass effects in
the QCD calculations of refs.\cite{ALTA,Buch:93} (in particular in the
decay $b \to c\bar c s$ (see also \cite{Voloshin} )) and taking into account
various renormalization
scale uncertainties improves the situation considerably.
Bagan et al. find \cite{Bagan}:
$B_{SL}=(12.0\pm 1.4)\%$ and $\bar B_{SL}=(11.2\pm1.7)\%$
for the pole quark masses and $\overline{MS}$ masses respectively.
Within existing uncertainties, this result does not disagree
significantly with the experimental value, although it is still
somewhat on the high side.
\subsection{$\Delta S=2$ and $\Delta B=2 $ Transitions}
The $M_{12}$ amplitude describing the $K^{0}-\bar K^{0}$ mixing is
given as follows
\begin{equation}
M_{12}(\Delta S=2)=\frac{G_F^2}{12\pi^2}F_K^2 B_K m_K M_W^2
\left[\lambda_c^{*2}\eta_1 S(x_c) +\lambda_t^{*2}\eta_2 S(x_t)
+2\lambda_c^{*}\lambda_t^{*}\eta_3 S(x_c,x_t)\right]
\end{equation}
with $x_i=m_i^2/M_W^2$, $\lambda_i=V_{id}V_{is}^{*}$,
 $S(x_i)$ denoting the Inami-Lim functions
resulting from box diagrams and $\eta_i$ representing QCD corrections.
The parameter $B_K$ is defined in (\ref{bk}).
The corresponding amplitude for the $B_d^{o}-\bar B_d^{o}$ mixing is
dominated by the box diagrams with top quark exchanges and given by
\begin{equation}
\mid M_{12}(\Delta B=2)\mid =\frac{G_F^2}{12\pi^2}F_B^2 B_B m_B M_W^2
\mid V_{td} \mid^2 \eta_B S(x_t)
\end{equation}
where we have set $V_{tb}=1$. A similar formula exists for
$B_s^{o}-\bar B_s^{o}$.
In the leading order
$\eta_i$ are given
roughly
\cite{GIL,BBH,KSY,Flynn,DFP} as follows:
$\eta_1=0.85,~ \eta_2=0.62,~ \eta_3=0.36,~\eta_B=0.60.$
As of 1995 the coefficients $\eta_i$ and $\eta_B$ are known including
NLO corrections \cite{BJW,HNa,HNb}.
It has been stressed in these papers that the LO results for $\eta_i$
suffer from sizable scale uncertainties, as large
as $\pm 20\%$ for $\eta_1$ and $\pm 10\% $ for the remaining $\eta_i$.
As demonstrated in \cite{BJW,HNa,HNb} these uncertainties
are considerably reduced in the products like
$\eta_1 S(x_c),~\eta_2 S(x_t),~\eta_3 S(x_c,x_t)$ and $\eta_B S(x_t)$
  provided NLO corrections
are taken into account. For $m_c=\bar m_c(m_c)=1.3\pm0.1~GeV$ and
$m_t=\bar m_t(m_t)=170\pm 15~GeV$
one finds:
\begin{equation}\label{KNLO}
\eta_1=1.3\pm 0.2\qquad \eta_2=0.57\pm0.01\qquad \eta_3=0.**\pm0.04
\qquad
\eta_B=0.55\pm0.01
\end{equation}
where the "**" in $\eta_3$ will be public soon \cite{HNb}.
It should be stressed that $\eta_i$ given here are so defined that
the relevant $B_K$ and $B_B$ non-perturbative factors (see (\ref{bk}))
are renormalization group invariant.
Let us list the main implications of these results:
\begin{itemize}
\item
The enhancement of $\eta_1$ implies the enhacement of the short
distance contribution to the $K_L-K_S$ mass difference so that for
$B_K=3/4$ as much as $80\%$ of the experimental value can be
attributed to this contribution \cite{HNa}.
\item
The improved calculations of $\eta_2$ and $\eta_3$
combined with the analysis of the CP violating parameter $\varepsilon_K$
allow an improved determination of the parameters $\eta$ and $\varrho$
in the CKM matrix \cite{BLO,HNb}.
\item
Similarly the improved calculation of $\eta_B$ combined with the
analysis of $B^0_d-\bar B^0_d$ mixing allows an improved determination
of the element $\mid V_{td}\mid$ \cite{BLO}:
\begin{equation}
\mid V_{td} \mid=
8.7\cdot 10^{-3}\left [
\frac{200~MeV}{\sqrt{B_B}F_B}\right ]
\left [\frac{170~GeV}{\bar m_t(m_t)} \right ]^{0.76}
\left [\frac{x_d}{0.72} \right ]^{0.5}
\left [\frac{1.50~ps}{\tau_B} \right ]^{0.5}
\end{equation}
This using all uncertainties (see below) gives:
\begin{equation}\label{vtd}
\mid V_{td} \mid = (9.6\pm 3.0)\cdot 10^{-3}
\quad
=>\quad (9.3 \pm 2.5 )\cdot 10^{-3}
\end{equation}
with the last number obtained after the inclusion of the
$\varepsilon_K$-analysis \cite{BLO}.
\end{itemize}
Concerning the parameter $B_K$, the most recent analyses
using the lattice methods
\cite{SH0,Ishizuka} ($B_K=0.83\pm 0.03$) and the $1/N$ approach
 of \cite{BBG0}
modified somewhat in \cite{Bijnens} give results in the ball park
of the $1/N$ result $B_K=0.70\pm 0.10$ obtained long
time ago \cite{BBG0}. In particular the analysis of Bijnens and Prades
\cite{Bijnens} seems to have explained the difference between these values
for $B_K$ and the lower values obtained using the QCD Hadronic Duality
approach \cite{Prades} ($B_K=0.39\pm 0.10$) or using SU(3) symmetry and
 PCAC
($B_K=1/3$) \cite{Donoghue}. This is gratifying because such low values for
$B_K$ would require $m_t>250~GeV$ in order to explain the experimental
value of $\varepsilon$ \cite{AB,BLO,HNb}.

There is a vast literature on the lattice calculations of $F_B$.
Based on a review by Chris Sachrajda \cite{Chris}, the recent extensive
study by Duncan et al. \cite{Duncan} and the analyses in \cite{Latt}
we conclude:
$F_{B_d}=(180\pm40)~MeV$. This together with the earlier result of
the European Collaboration for $B_B$, gives
$F_{B_d}\sqrt{B_{B_d}}=195\pm 45~MeV$.
The reduction of the error in this important quantity is desirable.
These results for $F_B$ are compatible with the results obtained using
QCD sum rules (eg.\cite{QCDS}). An interesting upper bound
$F_{B_d}<195~MeV$ using QCD dispersion relations has also recently
been obtained \cite{BGL}.
\subsection{$\Delta S=1$ Hamiltonian and $\varepsilon'/\varepsilon$}
The effective Hamiltonian for $\Delta S=1$ transitions is given
as follows:
\begin{equation}\label{dels1}
{\cal H}_{eff}(\Delta S=1) = \f{G_F}{\sqrt{2}} V_{us}^* V_{ud}
 \sum_{i=1}^{10} \left[ z_i(\mu)+\tau y_i(\mu)\right] Q_i
\end{equation}
where $\tau=-(V_{td}V_{ts}^*)/(V_{ud}V_{us}^*)$.
The coefficients of all ten operators are known
including NLO QCD and QED effects in NDR and HV schemes due to
the independent work of Munich and Rome groups
\cite{BJLW1,BJLW2,BJLW,ROMA1,ROMA2}.
A direct application of these
results is the calculation of
Re($\varepsilon'/\varepsilon$) which measures the ratio of direct
to indirect
CP violation in $K\to\pi\pi$ decays. In the standard model
$\varepsilon'/\varepsilon $ is governed by QCD penguins and
electroweak (EW)
penguins \cite{GIL0}.
With increasing $m_t$
the EW-penguins become increasingly important \cite{FLYNN,BBH} and entering
$\varepsilon'/\varepsilon$ with the opposite sign to QCD-penguins suppress
this ratio for large $m_t$. For $m_t\approx 200~GeV$ the ratio can even
be zero \cite{BBH}.
This strong cancellations between these two contributions was one of
the prime motivations for the NLO calculations performed in Munich
and Rome. Although these calculations can be regarded as an important
step towards a reliable theoretical prediction for
$\varepsilon'/\varepsilon$ the situation is clearly not satisfactory
at present.
Indeed $\varepsilon'/\varepsilon$ is plagued with uncertainties related to
non-perturbative B-factors which multiply $m_t$ dependent functions in a
formula like (\ref{PBEE}). Several of these B-factors can be extracted from
leading
CP-conserving $K\to\pi\pi$ decays \cite{BJLW}. Two important B-factors
($B_6=$ the dominant QCD penguin ($Q_6$) and $B_8=$ the dominant electroweak
 penguin ($Q_8$))
cannot be determined this way and one has to use lattice or $1/N$ methods
to predict Re($\varepsilon'/\varepsilon$).

 An analytic formula for
Re($\varepsilon'/\varepsilon$) as a function of
$m_t,~\Lambda_{\overline{MS}},~B_6,~B_8,~m_s$ and $V_{CKM}$ can be found
in \cite{BLAU}. A very simplified version of this formula is given as follows
\begin{equation}\label{7e}
{\rm Re}(\frac{\varepsilon'}{\varepsilon})=12\cdot 10^{-4}\left [
\frac{\eta\lambda^5 A^2}{1.7\cdot 10^{-4}}\right ]
\left [\frac{150~MeV}{\bar m_s(m_c)} \right ]^2
\left [\frac{\Lms^{(4)}}{300~MeV} \right ]^{0.8}
[B_6-Z(x_t)B_8]
\end{equation}
where $Z(x_t)= 0.175\cdot x_t^{0.93}.$  Note the strong dependence on
$\Lms$ pointed out in \cite{BJLW}.
For $m_t=170\pm13~GeV$, $\bar m_s(m_c)\approx 150\pm20~MeV$ \cite{Jamin}
and using $\varepsilon_K$-analysis to determine
$\eta$ one finds using the formulae in \cite{BJLW,BLAU} roughly
\begin{equation}\label{8}
-1\cdot 10^{-4} \leq Re(\frac{\varepsilon'}{\varepsilon})\leq 15\cdot 10^{-4}
\end{equation}
if $B_6=1.0\pm 0.2$ and $B_8=1.0\pm 0.2$ are used.
Such values are found in the $1/N$
approach \cite{BBG} and using lattice methods: \cite{SH1}
and \cite{SH1,SH2} for $B_6$ and $B_8$ respectively.
A very recent analysis of the Rome group \cite{ROMA3} gives a smaller
range,
 $Re(\varepsilon'/\varepsilon)=(3.1\pm 2.5)\cdot 10^{-4}$, which is
 however compatible
with (\ref{8}). Similar results are found with hadronic matrix elements
calculated in the chiral quark model \cite{Stefano}.
However $\varepsilon'/\varepsilon$ obtained in \cite{DORT} is
substantially larger and about $2 \cdot 10^{-3}$.

The experimental situation on Re($\varepsilon'/\varepsilon$) is unclear
at present.
 While
the result of NA31 collaboration at CERN with Re$(\varepsilon'/\varepsilon)
= (23 \pm 7)\cdot 10^{-4}$ \cite{WAGNER} clearly indicates
direct CP violation, the value of E731 at Fermilab,
Re$(\varepsilon'/\varepsilon) = (7.4 \pm 5.9)\cdot 10^{-4}$
\cite{GIBBONS} is compatible with superweak theories \cite{WO1} in which
$\varepsilon'/\varepsilon = 0$.
The E731 result is in the ball park of the theoretical estimates.
The NA31 value appears a bit high compared to the range given in
(\ref{8}).

 Hopefully, in about five years the
experimental situation concerning $\varepsilon'/\varepsilon$ will be
clarified through the improved measurements by the two collaborations
at the $10^{-4}$ level and by experiments at the $\Phi$ factory in
Frascati.
One should also hope that the theoretical situation of
$\varepsilon'/\varepsilon$ will improve by then to confront the new data.
\subsection{$\Delta B=1$ Effective Hamiltonian}
The effective hamiltonian for $\Delta B=1$ transitions involving
operators $Q_1,..Q_{10}$ (with corresponding changes of flavours)
is also known including NLO corrections \cite{BJLW}. It has been
used in the study of CP asymmetries in B-decays \cite{Fleischer:94}.
\subsection{$K\to\pi^o e^+e^-$}
The effective Hamiltonian for $K\to\pi^0 e^+e^-$  is given
as follows:
\begin{equation}\label{dels}
{\cal H}_{eff}(K\to\pi^0 e^+e^-) = \f{G_F}{\sqrt{2}} V_{us}^* V_{ud}
 \left[\sum_{i=1}^{6,9V} \left[ z_i(\mu)+\tau y_i(\mu)\right] Q_i
+\tau y_{10A}(M_W)Q_{10A}\right]
\end{equation}
where $Q_{9V}$ and $Q_{10A}$ are given by (\ref{9V})
 with $\bar b s$ replaced by $\bar s d $.

 Whereas in $K \to \pi \pi$ decays the CP violating
contribution is a tiny part of the full amplitude and the direct CP
violation is expected to be at least by three orders of magnitude
smaller than the indirect CP violation, the corresponding hierarchies
are very different for the rare decay $K_L\to\pi^o e^+e^-$ .
At lowest order in
electroweak interactions
this decay takes place only if CP symmetry is
violated \cite{GIL1}.
Moreover, the direct CP violating contribution is predicted to be
larger than
the indirect one. The CP conserving contribution to the amplitude
comes from a two photon exchange.
  The studies in
 \cite{Seghal,PICH} indicate that it is
 smaller than the direct CP
violating contribution.

The size of the indirect CP violating contribution will be
known once the CP conserving decay $K_S \to \pi^0 e^+ e^-$ has been
measured \cite{BARR}. On the other hand the direct CP violating
contribution can
be fully calculated as a function of $m_t$, CKM parameters and the
QCD coupling constant $\alpha_s$. There are practically no theoretical
uncertainties related to hadronic matrix elements in this part,
because the relevant matrix elements of the operators $Q_{9V}$ and
$Q_{10A}$ can be extracted from the well-measured decay
$K^+\to \pi^0 e^+ \nu$.
The NLO QCD corrections to the direct CP violating part
have been calculated in \cite{BLMM}
reducing certain ambiguities present in
leading order analyses \cite{GIL2} and enhancing somewhat
the theoretical prediction.
For $m_t=170\pm 10~GeV$ one finds \cite{BLMM}
\begin{equation}\label{12}
Br(K_L\to\pi^0 e^+ e^-)_{dir}=(5.\pm 2.)\cdot 10^{-12}
\end{equation}
where the error comes dominantly from the uncertainties in the CKM
parameters.
This should be compared with the present estimates of the other two
contributions:  $Br(K_L\to\pi^o e^+e^-)_{indir}\leq 1.6\cdot 10^{-12}$
and $Br(K_L\to\pi^o e^+e^-)_{cons}\approx(0.3-1.8)\cdot 10^{-12}$ for
the indirect
CP violating and the CP conserving contributions respectively \cite{PICH}.
Thus direct
CP violation is expected to dominate this decay.
The present experimental
bounds
\begin{equation}
Br(K_L\to\pi^0 e^+ e^-) \leq\left\{ \begin{array}{ll}
4.3 \cdot 10^{-9} & \cite{harris} \\
5.5 \cdot 10^{-9} & \cite{ohl} \end{array} \right.
\end{equation}
are still by three orders of magnitude away from the theoretical
expectations in the Standard Model. Yet the prospects of getting the
required sensitivity of order $10^{-11}$--$10^{-12}$ in five years are
encouraging \cite{CPRARE}.
\subsection {$B\to X_s\gamma$}
The effective hamiltonian for $B\to X_s\gamma$ at scales $\mu=O(m_b)$
is given by
\be \label{Heff_at_mu}
{\cal H}_{eff}(b\to s\gamma) = - \f{G_F}{\sqrt{2}} V_{ts}^* V_{tb}
\left[ \sum_{i=1}^6 C_i(\mu) Q_i + C_{7\gamma}(\mu) Q_{7\gamma}
+C_{8G}(\mu) Q_{8G} \right]
\ee
 The perturbative QCD effects are very important in this decay.
They are known
\cite{Bert,Desh} to enhance $B\to X_s\gamma$ in
the SM by 2--3
times, depending on the top quark mass. Since the first analyses
in \cite{Bert,Desh} a lot of progress has been made in calculating
the QCD effects begining with the work in \cite{Grin,Odon}.

A peculiar feature of the renormalization group analysis
in $B\to X_s\gamma$ is that the mixing under infinite renormalization
between
the set $(Q_1...Q_6)$ and the operators $(Q_{7\gamma},Q_{8G})$ vanishes
at the one-loop level. Consequently in order to calculate
the coefficients
$C_{7\gamma}(\mu)$ and $C_{8G}(\mu)$ in the leading logarithmic
approximation, two-loop calculations of ${\cal{O}}(e g^2_s)$
and ${\cal{O}}(g^3_s)$
are necessary. The corresponding NLO analysis requires the evaluation
of the mixing in question at the three-loop level.

At present, the coefficients $C_{7\gamma}$ and $C_{8G}$ are only known
in the leading logarithmic approximation.
However the peculiar feature of this decay mentioned above caused
that the first fully correct calculation of the leading  anomalous
dimension matrix has been obtained only in 1993 \cite{CFMRS:93,CFRS:94}.
It has been
confirmed subsequently in \cite{CCRV:94a,CCRV:94b,Mis:94}.
The NLO corrections are only partially known.
The two-loop mixing
involving the operators
$Q_1.....Q_6$ is the same as in section 4.5. The two-loop mixing
in the sector $(Q_{7\gamma},Q_{8G})$ has been calculated
last year \cite{MisMu:94}.
The three loop mixing between
the set $(Q_1...Q_6)$ and the operators $(Q_{7\gamma},Q_{8G})$
 has not be done. The $O(\alpha_s)$
corrections to $C_{7\gamma}(M_W)$ and $C_{8G}(M_W)$ have been considered
in \cite{Yao1}. Gluon corrections to the matrix elements of magnetic
penguin operators have been calculated in \cite{AG1,AG2}.

The leading
logarithmic calculations of $Br(B \ra X_s \gamma)$
\cite{Grin,CFRS:94,CCRV:94a,Mis:94,AG1,BMMP:94}
 are based on the
spectator model corrected for short-distance QCD effects discussed
above.
As we have stressed previously support for this approximation
comes from the $1/m_b $ expansions.
A critical analysis of theoretical and
experimental
uncertainties present in the LO prediction for Br(\Bsg)
has been made in \cite{BMMP:94} giving
\be
Br(B \ra X_s\gamma)_{TH} = (2.8 \pm 0.8) \times 10^{-4}.
\label{theo}
\ee
where the error is dominated by the uncertainty in
choice of the renormalization scale
$m_b/2<\mu<2 m_b$ as first stressed by Ali and Greub \cite{AG1} and confirmed
in \cite{BMMP:94}.
	Since \Bsg is dominated by QCD effects, it is not surprising
that this scale-uncertainty in the leading order
is particularly large.

The \Bsg decay has already been measured.
In 1993
CLEO reported \cite{CLEO}
$Br(B \ra K^* \gamma) = (4.5 \pm 1.5 \pm 0.9) \times 10^{-5}.$
In 1994 first measurement of the inclusive rate has been
presented by CLEO \cite{CLEO2}:
\be
Br(B \ra X_s\gamma) = (2.32 \pm 0.57 \pm 0.35) \times 10^{-4}.
\label{incl}
\ee
where the first error is statistical and the second is systematic.
This result agrees with (\ref{theo}) very well although
the theoretical and experimental errors should be decreased in
the future in order to reach a definite conclusion and to see
whether some contributions beyond the standard model
are required. In any case the agreement of the
theory with data is consistent with the large QCD enhancement
of \Bsg. Without this enhancement the theoretical prediction
would be at least by a factor of 2 below the data.
The partial inclusion of NLO corrections done in
\cite{Ciu:94}
lowers the theoretical branching ratio down to
$Br(B\to X_s\gamma)=(1.9\pm 0.2\pm 0.5)\cdot 10^{-4}$,
We have to wait however for the final complete NLO calculation
which should considerably reduce theoretical uncertainties in the
leading order as formally demonstrated in \cite{BMMP:94}.
\subsection{$B\to X_s e^+e^-$ Beyond Leading Logarithms}
The effective hamiltonian for $B\to X_s e^+e^-$ at scales $\mu=O(m_b)$
is given by
\be \label{Heff2_at_mu}
{\cal H}_{eff}(b\to s e^+e^-) =
{\cal H}_{eff}(b\to s\gamma)  - \f{G_F}{\sqrt{2}} V_{ts}^* V_{tb}
\left[ C_{9V}(\mu) Q_{9V}+
C_{10A}(M_W) Q_{10A}    \right]
\ee
where
${\cal H}_{eff}(b\to s\gamma)$ is given in (\ref{Heff_at_mu}).
In addition to the operators relevant for $B\to X_s\gamma$,
there are two new operators
$Q_{9V}$ and $Q_{10A}$
which appeared already in the decay \kpiee
except for an appropriate change of quark flavours
and the fact that now $\mu={\cal O}(m_b)$ instead of
$\mu={\cal O}(1~GeV)$ should be considered. There is a large literature
on this dacay. In particular Hou et al \cite{HWS:87} stressed
the strong dependence of $B\to X_s e^+e^-$ on $m_t$.
Further references to phenomenology can be found in \cite{BuMu:94}.

The QCD corrections to this decay have been calculated
 over the last years with increasing precision by several
groups \cite{GSW:89,GDSN:89,CRV:91,Mis:94} culminating in two complete
next-to-leading QCD calculations
\cite{Mis:94,BuMu:94} which agree with each other.
An extensive numerical analysis of the differential decay rate
including NLO corrections has been presented in \cite{BuMu:94}.
The NLO corrections enhance the leading order results by roughly $15\%$.
The differential decay rate normalized to $\Gamma(B\to X_c e\bar\nu)$,
varies for
$0.1 <(p_{e^+}+p_{e^-})^2/m_b^2< 0.8$
between $1\cdot 10^{-4}$ and $1\cdot 10^{-5}$ when $m_t=170~GeV$ and
$\Lambda_{\overline{MS}}=225~MeV$ are chosen.
Similar result has been obtained by
Misiak \cite{Mis:94}. The $1/m^2_b$ corrections calculated in \cite{FALK}
enhance these results by roughly $10\%$.
\subsection{$K_L\to\pi^o\nu\bar\nu$, $K^+\to\pi^+\nu\bar\nu$,
$ K_L\to\mu\bar\mu$, $B\to\mu\bar\mu$ and $B\to X_s\nu\bar\nu$}
$K_L\to\pi^o\nu\bar\nu$, $K^+\to\pi^+\nu\bar\nu$,
$B\to\mu\bar\mu$ and $B\to X_s\nu\bar\nu$ are the theoretically
cleanest decays in the field of rare decays.
$K_L\to\pi^o\nu\bar\nu$,
$B\to\mu\bar\mu$ and $B\to X_s\nu\bar\nu$
 are dominated by short distance loop diagrams
involving the top quark.  $K^+\to\pi^+\nu\bar\nu$ receives
additional sizable contributions from internal charm exchanges.
The decay $K_L\to \mu\bar\mu$ receives substantial long distance
contributions and  consequently suffers from large theoretical
uncertainties. This is very unfortunate because with the existing data
this decay could offer a good determination of the parameter $\varrho$
in the CKM matrix.
The most accurate is the measurement from Brookhaven \cite{PRINZ}:
$Br(K_L\to \bar\mu\mu) = (6.86\pm0.37)\cdot 10^{-9}$,
which is somewhat lower than the KEK-137 result:
$(7.9\pm 0.6 \pm 0.3)\cdot 10^{-9}$ \cite{Akagi}.
For the short distance contribution I find using the formulae of
\cite{BB3}:
$Br(K_L\to \bar\mu\mu)_{SD} = (1.5\pm 0.8)\cdot 10^{-9}$.
Details on this decay can be found in \cite{PRINZ,BB3}.
More promising from theoretical point of view is the parity-violating
asymmetry in $K^+\to \pi^+\mu^+\mu^-$ \cite{GENG,BB5}.

The NLO QCD corrections to all these decays
have been calculated in a series of papers by Buchalla and
myself \cite{BB1,BB2,BB3,BB5}. These calculations
considerably reduced the theoretical uncertainties
due to the choice of the renormalization scales present in the
leading order expressions \cite{DDG}. Since the relevant hadronic matrix
elements of the weak currents entering $K\to \pi\nu\bar\nu$
can be measured in the leading
decay $K^+ \rightarrow \pi^0 e^+ \nu$, the resulting theoretical
expressions for Br( $K_L\to\pi^o\nu\bar\nu$) and Br($K^+\to\pi^+\nu\bar\nu$)
  are
only functions of the CKM parameters, the QCD scale
 $\Lambda \overline{_{MS}}$
 and the
quark masses $m_t$ and $m_c$.
The long distance contributions to
$K \rightarrow \pi \nu \bar{\nu}$ have been
considered in \cite{RS} and found to be very small.
Similar comments apply to $B\to\mu\bar\mu$ and
$B\to X_s\nu\bar\nu$ except that $B\to\mu\bar\mu$ depends on the
B-meson decay constant $F_B$ which brings in the main theoretical
uncertainty.

The explicit expressions for $Br(\kpnn)$ and $Br(\klpnn)$ are given as
 follows
\be\label{bkpn}
Br(\kpn)=4.57\cdot 10^{-11} A^4 X^2(x_t)
\cdot\left[ \eta^2+(\varrho_0-\varrho)^2 \right]            \ee
\be\label{bklpn}
Br(K_L\to\pi^0\nu\bar\nu)=1.91\cdot 10^{-10}\eta^2 A^4 X^2(x_t)
\end{equation}
Here
\begin{equation}
\varrho_0=1+\frac{P_0(K^+)}{A^2 X(x_t)}
\quad\quad
X(x_t) = 0.65\cdot x_t^{0.575}
\end{equation}
where the NLO correction calculated in \cite{BB2} is included
in $X(x_t)$ if $m_t\equiv\bar m_t(m_t)$.
 Next $P_0(K^+)=0.40\pm0.09$ \cite{BB3,BB4} is a function of $m_c$ and
 $\Lambda_{\overline{MS}}$ and includes the residual uncertainty
due to the renormalization scale $\mu$. The absence of $P_0$ in
(\ref{bklpn}) makes $\klpnn$ theoretically even cleaner than $\kpnn$.
We should remark that (\ref{bkpn}) is an approximation. A more accurate
formula is given in \cite{BB3}.

Similarly for $B_s\to \mu\bar\mu$ one has \cite{BB2}
\begin{equation}
Br(B_s\to \mu\bar\mu)=
4.1\cdot 10^{-9}\left [ \frac{F_{B_s}}{230~MeV}\right ]^2
\left [\frac{\bar m_t(m_t)}{170~GeV} \right ]^{3.12}
\left [\frac{\mid V_{ts}\mid}{0.040} \right ]^2
\left [\frac{\tau_{B_s}}{1.6 ps} \right ]
\end{equation}

The impact of NLO calculations is best illustrated by giving the
scale uncertainties in the leading order and after the inclusion
of the next-to-leading corrections:
\begin{equation}
Br(\kpn)=(1.00\pm0.20)\cdot 10^{-10}\quad =>\quad
(1.00\pm0.05)\cdot 10^{-10}
\end{equation}
\begin{equation}
Br(\klpnn)=(3.00\pm0.30)\cdot 10^{-11}\quad =>\quad
(3.00\pm0.04)\cdot 10^{-11}
\end{equation}
\begin{equation}
Br(B_s\to \mu\bar\mu)=(4.10\pm0.50)\cdot 10^{-9}\quad =>\quad
(4.10\pm0.05)\cdot 10^{-9}
\end{equation}
The reduction of the scale uncertainties is truly impressive.

The present experimental bound on $Br(K^+\to \pi^+\nu\bar\nu)$
is $5.2 \cdot 10^{-9}$ \cite{Atiya} (a preliminary result from this
group is $3.0 \cdot 10^{-9}$). An improvement by one order
of magnitude is expected at AGS in Brookhaven for the coming years.
The present upper bound on $Br(K_L\to \pi^0\nu\bar\nu)$ from
Fermilab experiment E799I is $5.8 \cdot 10^{-5}$ \cite{E799}.
FNAL-E799II
 expects to reach
the accuracy ${\cal O}(10^{-8})$ and the future experiments at FNAL
and KEK will hopefully be able to reach the standard model
expectations. The latter are given for both decays at present as follows:
\begin{equation}
Br(\kpn)=(1.1\pm 0.4)\cdot 10^{-10}\quad,\quad
Br(\klpnn)=(3.0\pm 2.0)\cdot 10^{-11}
\end{equation}
\section{Finalists}
\subsection{General Remarks}
{}From tree level K decays sensitive to $V_{us}$ and tree level B decays
 sensitive to $V_{cb}$ and $V_{ub}$ we have:
\begin{equation}\label{2}
\lambda=0.2205\pm0.0018
\qquad
\mid V_{cb} \mid=0.041\pm0.003\quad =>\quad A=0.85\pm 0.06
\end{equation}
\begin{equation}\label{2.94}
\left| \frac{V_{ub}}{V_{cb}} \right|=0.08\pm0.03
\quad => \quad
\sqrt{\varrho^2+\eta^2} =0.36\pm0.14
\end{equation}
The main recent progress here, is the improved determination of
$\mid V_{cb}\mid$ due to experimental \cite{VCB} and theoretical
efforts \cite{Neubert,Bigi,Brown}. Although some further reduction
of the errors could be expected in the future,
it is difficult to imagine at present that
in tree level B-decays
a better accuracy than $\Delta\vcb=\pm 2\cdot 10^{-3}$ and
$\Delta\vub=\pm 0.01$ ($\Delta R_b=\pm 0.04$) could be achieved unless some
dramatic improvements in the theory and experiment will take place.
It is therefore of interest to look simultaneously at other decays in
order to improve the determination of these parameters. For instance
as stressed in \cite{AB94,AB94A}, it is in principle possible  to determine
all CKM parameters without any hadronic uncertainties
  although this will
require heroic experimental efforts.
Indeed using the loop induced decays or transitions
which are fully governed by short distance
physics {\it simultaneously} with CP asymmetries in B-decays
clean and
precise determinations of $\vcb$, $\vub$, $\vtd$, $\varrho$ and $\eta$
can be achieved.
In this respect the most promising from the
theoretical point of view are the following four:
i) CP-Asymmetries in $B^o$-Decays,
ii) $K_L\to\pi^o\nu\bar\nu$,
iii) $\kpnn$ and iv) $(B^o_d-\bar B_d^o)/(B^o_s-\bar B_s^o)$.
Let us summarize their main virtues one-by-one.
\subsection{CP-Asymmetries in $B^o$-Decays}
The CP-asymmetry in the decay $B_d^\circ \rightarrow \psi K_S$ allows
 in the standard model
a direct measurement of the angle $\beta$ in the unitarity triangle
without any theoretical uncertainties. This has been first pointed out
by Bigi and Sanda \cite {BSANDA}, analyzed in detail already in \cite{BSS}
and during the past years
discussed by many authors \cite{NQ}.
 Similarly the decay
$B_d^\circ \rightarrow \pi^+ \pi^-$ gives the angle $\alpha$, although
 in this case strategies involving
other channels are necessary in order to remove hadronic
uncertainties related to penguin contributions
\cite{CPASYM}.
The determination of the angle~$\gamma$ from CP asymmetries in neutral
B-decays is more difficult but not impossible
\cite{RF}. Also charged B decays could be useful in
this respect \cite{Wyler}.

Since in the usual unitarity triangle  one side is known,
it suffices to measure
two angles to determine the triangle completely. This means for instance
that the measurements of $\sin 2\beta$ and $\sin 2\alpha$
through the asymmetries  $ A_{CP}(\psi K_S)$ and $ A_{CP}(\pi^+\pi^-)$
can determine
the parameters $\varrho$ and $\eta$.
The main virtues of this determination are as follows:
\begin{itemize}
\item No hadronic or $\Lambda_{\overline{MS}}$ uncertainties.
\item No dependence on $m_t$ and $V_{cb}$ (or A).
\end{itemize}
As various analyses \cite{BLO,ALI,ROMA3} of the unitarity triangle
show, $\sin(2\beta)$ is expected to be large:
$\sin(2\beta)\approx 0.6\pm 0.2$.
The predictions for $\sin(2\gamma)$ and $\sin(2\alpha)$ are very
uncertain on the other hand.

\subsection{$K_L\to\pi^o\nu\bar\nu$}
As we have discussed above $K_L\to\pi^o\nu\bar\nu$ is the theoretically
cleanest decay in the field of rare K-decays.
Moreover it proceeds almost entirely through direct CP violation \cite{LI}.
 The main features of this decay are:
\begin{itemize}
\item No hadronic uncertainties
\item  $\Lambda_{\overline{MS}}$ and renormalization scale uncertainties
at most $\pm 1\%$ \cite{BB2}.
\item Strong dependence on $m_t$ and $V_{cb}$ (or A).
\end{itemize}
\subsection{$\kpnn$}
$K^+\to\pi^+\nu\bar\nu$ is CP conserving and receives
contributions from
both internal top and charm exchanges.
$K^+\to\pi^+\nu\bar\nu$ is the second best decay in the field
of rare decays.
The main features of this decay are:
\begin{itemize}
\item Hadronic uncertainties below $1\%$ \cite{RS}
\item  $\Lambda_{\overline{MS}}$, $m_c$ and renormalization scales
 uncertainties at most $\pm (5-10)\%$ \cite{BB3}.
\item Strong dependence on $m_t$ and $V_{cb}$ (or A).
\end{itemize}
\subsection{$(B^o_d-\bar B_d^o)/(B^o_s-\bar B_s^o)$ }
Measurement of $B^o_d-\bar B^o_d$ mixing parametrized by $x_d$ together
with  $B^o_s-\bar B^o_s$ mixing parametrized by $x_s$ allows to
determine $R_t$:
\begin{equation}\label{107b}
R_t = \frac{1}{\sqrt{R_{ds}}} \sqrt{\frac{x_d}{x_s}} \frac{1}{\lambda}
\qquad
R_{ds} = \frac{\tau_{B_d}}{\tau_{B_s}} \cdot \frac{m_{B_d}}{m_{B_s}}
\left[ \frac{F_{B_d} \sqrt{B_{B_d}}}{F_{B_s} \sqrt{B_{B_s}}} \right]^2
\end{equation}
where $R_{ds}$ summarizes SU(3)--flavour breaking effects.
Note that $m_t$ and $V_{cb}$ dependences have been eliminated this way
 and $R_{ds}$
contains much smaller theoretical
uncertainties than the hadronic matrix elements in $x_d$ and $x_s$
separately.
Provided $x_d/x_s$ has been accurately measured a determination
of $R_t$ within $\pm 10\%$ should be possible. Indeed the most
recent lattice result \cite{Duncan} gives $F_{B_d}/F_{B_s}=1.22\pm0.04$.
It would be useful to know $B_{B_s}/B_{B_d}$ with a similar precision.
For $B_{B_s}=B_{B_d}$ I find $R_{ds}=0.62\pm 0.07$. Consequently
rescaling the results of \cite{BLO}, obtained for $R_{ds}=1$, the
range $12 < x_s < 39 $ follows. Such a large mixing will not be
easy to measure. The main features of $x_d/x_s$ are:
\begin{itemize}
\item No $\Lambda_{\overline{MS}}$, $m_t$ and $V_{cb}$ dependence.
\item Hadronic uncertainty in SU(3)--flavour breaking effects of
      roughly $\pm 10\%$.
\end{itemize}
\subsection{$\sin (2\beta)$ from $K\to \pi\nu\bar\nu$}
It has been pointed out in \cite{BH} that
measurements of $Br(\kpn)$ and $Br(\klpn)$ could determine the
unitarity triangle completely provided $m_t$ and $V_{cb}$ are known.
In view of the strong dependence of these branching ratios on
$m_t$ and $V_{cb}$ this determination is not precise however \cite{BB4}.
On the other hand it has been noticed \cite{BB4} that the $m_t$ and
$V_{cb}$ dependences drop out in the evaluation of $\sin(2\beta)$.
Consequently $\kpnn$ and $\klpnn$ offer a
clean
determination of $\sin (2\beta)$ which can be confronted with
the one possible in $B^0\to\psi K_S$ discussed above.
Any difference in these two determinations
would signal new physics.
Choosing
$Br(\kpn)=(1.0\pm 0.1)\cdot 10^{-10}$ and
$Br(\klpn)=(2.5\pm 0.25)\cdot 10^{-11}$,
one finds \cite{BB4}
\begin{equation}\label{26}
\sin(2 \beta)=0.60\pm 0.06 \pm 0.03 \pm 0.02
\end{equation}
where the first error is "experimental", the second represents the
uncertainty in $m_c$ and  $\Lambda_{\overline{MS}}$ and the last
is due to the residual renormalization scale uncertainties. This
determination of $\sin(2\beta)$ is competitive with the one
expected at the B-factories at the beginning of the next decade.
\subsection{Precise Determinations of the CKM Matrix}
Using the first two finalists and $\lambda=0.2205\pm 0.0018$
\cite{LR}
it is possible to determine all the parameters
of the CKM matrix without any hadronic uncertainties
\cite{AB94}.
As illustrative examples we consider in table 2 three scenarios.
The first four rows give the assumed input parameters and their
experimental errors which are expected in the next decade.
The remaining rows give the results for
selected parameters.
The experimental errors on $Br(\klpnn)$
 to be achieved in the next 15 years are most probably unrealistic,
but I show this exercise anyway in order to motivate this very
challenging enterprise.
\begin{table}
\begin{center}
\begin{tabular}{|c||c||c|c|c|}\hline
& Central &$I$&$II$&$III$\\ \hline
$\sin(2\alpha)$ & $0.40$ &$\pm 0.08$ &$\pm 0.04$ & $\pm 0.02 $\\ \hline
$\sin(2\beta)$ & $0.70$ &$\pm 0.06$ &$\pm 0.02$ & $\pm 0.01 $\\ \hline
$m_t$ & $170$ &$\pm 5$ &$\pm 3$ & $\pm 3 $\\ \hline
$10^{11} Br(K_L)$ & $3$ &$\pm 0.30$ &$\pm 0.15$ & $\pm 0.15 $\\ \hline\hline
$\varrho$ &$0.072$ &$\pm 0.040$&$\pm 0.016$ &$\pm 0.008$\\ \hline
$\eta$ &$0.389$ &$\pm 0.044$ &$\pm 0.016$&$\pm 0.008$ \\ \hline
$\mid V_{ub}/V_{cb}\mid$ &$0.087$ &$\pm 0.010$ &$\pm 0.003$&$\pm 0.002$
 \\ \hline
$\mid V_{cb}\mid/10^{-3}$ &$39.2$ &$\pm 3.9$ &$\pm 1.7$&$\pm 1.3$\\ \hline
$\mid V_{td}\mid/10^{-3}$ &$8.7$ &$\pm 0.9$ &$\pm 0.4$ &$\pm 0.3$ \\
 \hline \hline
$\mid V_{cb}\mid/10^{-3}$ &$41.2$ &$\pm 4.3$ &$\pm 3.0$&$\pm 2.8$\\ \hline
$\mid V_{td}\mid/10^{-3}$ &$9.1$ &$\pm 0.9$ &$\pm 0.6$ &$\pm 0.6$ \\
 \hline
\end{tabular}
\end{center}
\centerline{}
\caption{Determinations of various parameters in scenarios I-III }
\end{table}
Table 2 shows very clearly the potential of CP asymmetries
in B-decays and of $\klpnn$ in the determination of CKM parameters.
It should be stressed that this high accuracy is not only achieved
because of our assumptions about future experimental errors in the
scenarios considered, but also because $\sin(2\alpha)$ is a
very sensitive function of $\varrho$ and $\eta$ \cite{BLO},
$Br(\klpnn)$
depends strongly on $\mid V_{cb}\mid$ and most importantly because
of the clean character of the quantities considered.

This results should be compared with
the expectations from a "standard" analysis
of the unitarity triangle
which is based on $\varepsilon_K$, $x_d$,
$\vcb$ and $\vub$ with the last two extracted from tree level decays.
As a typical analysis \cite{BLO} shows, even with optimistic assumptions
about the theoretical and experimental errors it will be difficult to
achieve the accuracy better than $\Delta\varrho=\pm 0.15$ and
$\Delta\eta=\pm 0.05$ this way.

In the last two rows of table 2 we show the
results for $\mid V_{cb} \mid$ and $\mid V_{td}\mid$
obtained using
$Br(\kpnn)= (1.0\pm 0.1)\cdot 10^{-10}$ for the scenario I and
$Br(\kpnn)= (1.0\pm 0.05)\cdot 10^{-10}$ for scenarios II and III
in place of $Br(\klpnn)$ with all other input parameters unchanged.
We observe that
due to the uncertainties present in the charm contribution to
$\kpnn$, which was absent in $\klpnn$, the determinations of
 $\mid V_{cb}\mid$ and  $\mid V_{td}\mid$  are less accurate,
 but still very interesting. In particular the error on
$\mid V_{td}\mid$ is much smaller than the one given in
(\ref{vtd}).

An alternative strategy is to use the measured value of $R_t$ instead
of $\sin(2\alpha)$.
The result of this exercise is shown in table 3.
Again the last two rows give the results when $\klpnn$ is replaced
by $\kpnn$.

The consistency of the determinations presented in tables 2 and 3
will offer an important test of the standard model.
Of particular interest will also be the comparison of $\mid V_{cb}\mid$
determined as suggested here with the value of this CKM element extracted
from tree level semi-leptonic  B-decays.
 Since in contrast to
$\klpnn$ and $\kpnn$, the tree-level decays are to an excellent approximation
insensitive to any new physics contributions from very high energy scales,
the comparison of these two determinations of $\mid V_{cb}\mid$ would
be a good test of the standard model and of a possible physics
beyond it.

\begin{table}
\begin{center}
\begin{tabular}{|c||c||c|c|c|}\hline
& Central &$I$&$II$&$III$\\ \hline
$R_t$ & $1.00$ &$\pm 0.10$ &$\pm 0.05$ & $\pm 0.03 $\\ \hline
$\sin(2\beta)$ & $0.70$ &$\pm 0.06$ &$\pm 0.02$ & $\pm 0.01 $\\ \hline
$m_t$ & $170$ &$\pm 5$ &$\pm 3$ & $\pm 3 $\\ \hline
$10^{11} Br(K_L)$ & $3$ &$\pm 0.30$ &$\pm 0.15$ & $\pm 0.15 $\\ \hline\hline
$\varrho$ &$0.076$ &$\pm 0.111$&$\pm 0.053$ &$\pm 0.031$\\ \hline
$\eta$ &$0.388$ &$\pm 0.079$ &$\pm 0.033$&$\pm 0.019$ \\ \hline
$\mid V_{ub}/V_{cb}\mid$ &$0.087$ &$\pm 0.014$ &$\pm 0.005$&$\pm 0.003$
 \\ \hline
$\mid V_{cb}\mid/10^{-3}$ &$39.3$ &$\pm 5.7$ &$\pm 2.6$&$\pm 1.8$\\ \hline
$\mid V_{td}\mid/10^{-3}$ &$8.7$ &$\pm 1.2$ &$\pm 0.6$ &$\pm 0.4$ \\
 \hline \hline
$\mid V_{cb}\mid/10^{-3}$ &$41.3$ &$\pm 5.8$ &$\pm 3.7$&$\pm 3.3$\\ \hline
$\mid V_{td}\mid/10^{-3}$ &$9.1$ &$\pm 1.3$ &$\pm 0.8$ &$\pm 0.7$ \\
 \hline
\end{tabular}
\end{center}
\centerline{}
\caption{ As in table 2 but with $\sin(2\alpha)$ replaced by $R_t$.}
\end{table}
\section{Final Remarks}
In this  compact review we have concentrated on rare decays and
CP violation in the
standard model. The structure of rare decays and of CP violation in
extensions of the
standard model may deviate from this picture.
Consequently the situation in this field could turn out to be very different
from the one presented here.
However in order to distinguish the standard model predictions from
the predictions of its extensions it is essential that the
theoretical calculations reach acceptable precision. In this
context we have emphasized the importance of the QCD calculations in
rare and CP violating decays. During the recent years a considerable
progress has been made in this field through the computation of NLO
contributions to a large class of decays. This effort reduced considerably
the theoretical uncertainties in the relevant formulae and thereby improved
the determination of the CKM parameters to be achieved in future
experiments. At the same time it should be stressed that whereas the
theoretical status of QCD calculations for rare semileptonic decays like
$K \to \pi\nu\bar\nu$, $B\to \mu\bar\mu$, $B \to X_s e^+ e^-$
 is fully satisfactory and
the status of $B\to X_s\gamma$ should improve in the coming years, a lot
remains to be done in a large class of non-leptonic decays or transitions
where non-perturbative uncertainties remain sizable.

I would like to thank the organizers for inviting me to this symposium
and for their great hospitality.

\vfill\eject
\end{document}